# ELASTICITY AND RELAXATION PROPERTIES OF ORAL FLUID

A.N. Rozhkov

Ishlinsky Institute for Problems in Mechanics of the Russian Academy of Sciences (IPMech RAS), 101 (k.1) Prospekt Vernadskogo, 119526, Moscow, Russian Federation, e-mail: rozhkov@ipmnet.ru

**Abstract.** The research is aimed at creating a method for rheological testing of viscoelastic fluids, the droplets of which, when stretched, form thinning filaments, i.e. exhibit the property of spinning. The typical example of such fluid is oral fluid (or mixed saliva). Among other things, the rheological features of the oral fluid actively affect the mechanisms of transmission of infection by airborne droplets. In this work, the oral fluid was studied both from the point of view of a participant in the transmission of infection and as a model for studying the rheology of viscoelastic fluids. The method is based on video recording of the stretching of a drop of the test liquid between the legs of the tweezers and the subsequent spontaneous thinning of the formed capillary filament of the liquid. The process is controlled by the competition of forces of inertia, elasticity, capillarity. By analyzing the video recording, it is possible to trace the contribution of each factor and, within the framework of the Oldroyd/Maxwell rheological constitutive equation (rheological model), determine the numerical values of all model constants. The obtained rheological characteristics of the oral fluid make it possible to theoretically model the processes of the formation of drops of oral fluid during sneezing, coughing, talking, as well as the processes of collision of drops with protective masks, filters and other obstacles. In general, the proposed method of rheological testing is applicable for studying a wide class of viscoelastic fluids, including biological ones. Among the latter are bronchial sputum, synovial fluid, reproductive fluids, and others. The method is distinguished by the simplicity of the experiment and the use of elementary equipment.

# RHEOLOGY IN AIRBORNE TRANSMISSION OF INFECTION

Airborne transmission of infection is the main one not only for the current coronavirus, but also for many other infectious diseases, such as tonsillitis, diphtheria, measles, rubella, chickenpox, etc. In these diseases, pathogens enter the air from droplets of saliva or pulmonary mucus when coughing, sneezing, talking. Oral fluid (or mixed saliva) consists of the secretions of the salivary glands (the main component) and many micro-additives present in the oral cavity and respiratory system. Airborne transmission of infection consists of two stages. First, when breathing, coughing, sneezing, an air enters the surrounding space from the mouth and lungs of the person. When air moves through the oral cavity, the liquid located here is captured and involved in movement. As a result, the outgoing air contains many tiny droplets of oral fluid, which, in the event of a person's illness, may contain pathogenic agents. In the second stage, other persons, inhaling the air with infected droplets, can bring the infection into their body and become infected.

To prevent the entry of drops from a sick person into the surrounding atmosphere, as well as to protect against the penetration of drops into the body of a healthy person, various filters and masks are used, the task of which is to delay the movement of drops. When sneezing, a large amount (up to 50,000) drops of mucus and saliva are released. The droplet velocity can be very high up to 40 m/s (different authors give values of 4.5 - 100 m/s), the droplets fly up to a distance of 4 - 6 m. The fate of the ejected drops can be roughly described as follows: drops with a diameter of more than 5 microns quickly settle under the action of gravity, and droplets less than 5 microns can remain in the air for a long time and be transported by air currents. When a relatively large droplet of liquid hits a hard surface or hits the material of a mask or filter, it can breakup into several smaller droplets, thus creating a new source of dangerous infection in the form of a new array of droplets.

As previously established, the formation of a splash and disintegration of a liquid droplet upon collision with an obstacle depends on its rheological (bulk and surface) properties [29]. In the case of collision of water droplets with an obstacle at high values of the Reynolds and Weber impact numbers ($Re_i = \rho v_i d_i / \mu$, $We_i = \rho v_i^2 d_i / \gamma$, $\rho$ is the density of the liquid, $v_i$ is the droplet velocity before collision, $d_i$ is the droplet diameter, $\mu$ is the viscosity liquid, $\gamma$ is the surface tension of the liquid), the process is controlled exclusively by inertia and surface tension [29, 40]. Experiments have shown that a drop upon collision retains its continuity at low Weber numbers and disintegrates at high Weber numbers. The transition from collision with preservation of continuity to the collision with droplet disintegration occurs in the range of Weber numbers $We_i^* \in (137, 206)$ (Fig. 1, a, b) [16, 30].

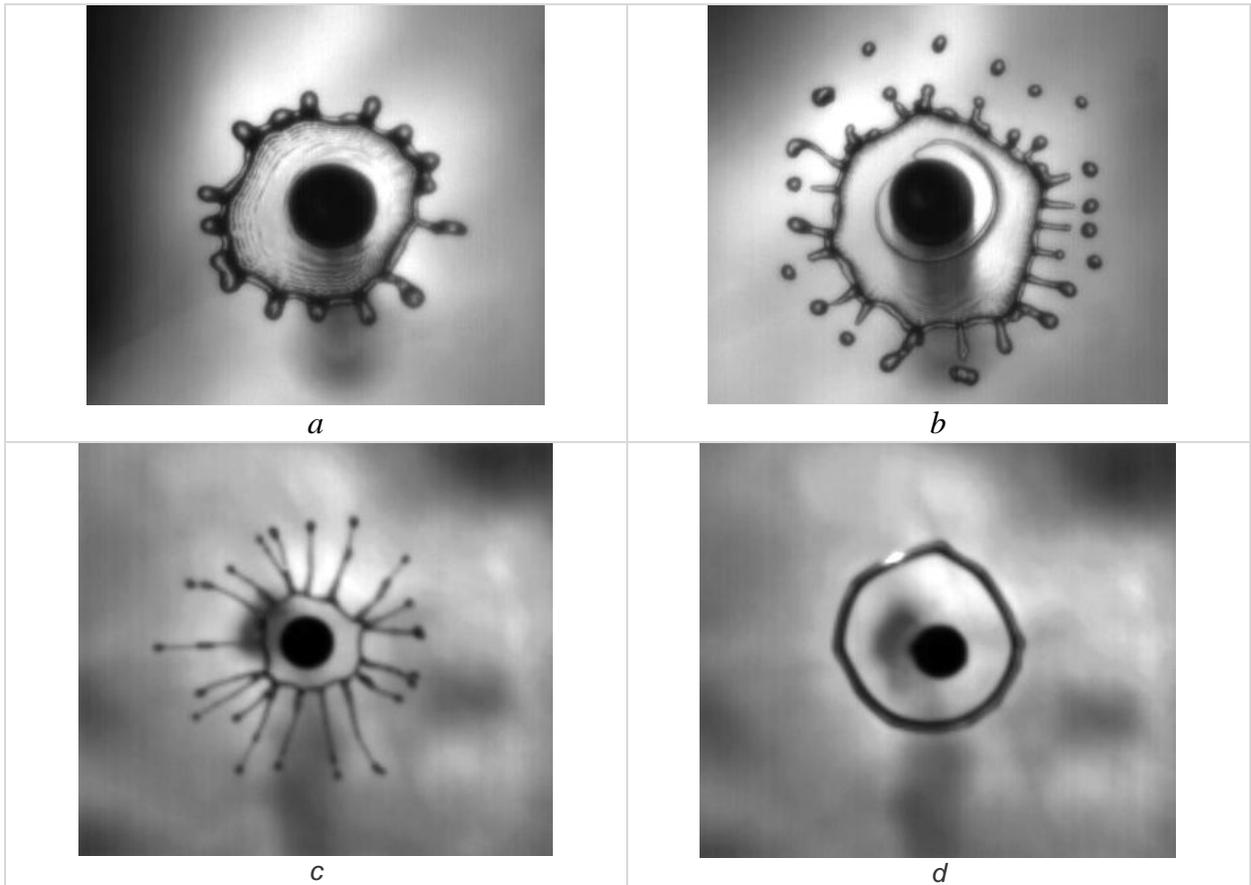

Fig. 1. Top view observations of the disintegration of the liquid drops when falling onto a disc-shaped obstacle with a diameter of 4 mm (looks like a black circle in the center of the frame) [29, 30]. 6 ms elapsed after the beginning of the collision. *a*, *b* - splashes of a 2.8 mm diameter water drop falling from a height of 18 and 27 cm ($We_i$=137 and 206), respectively. *c*, *d* - similar splashes of the drops of solutions of high molecular weight polyethylene oxide and polyacrylamide, respectively, with a concentration of 100 ppm. Drop falling height is 65 cm.

Studies [1, 14, 15, 20, 22, 44] have revealed the elasticity of the oral fluid, while its shear viscosity is only 1–4 times higher than the viscosity of water [26, 39]. The elasticity of the oral fluid is evidenced by the possibility to form stable thinning filament when stretched. The lifetime of such filament is measured in seconds, while a filament formed by water disintegrates during the parts of a millisecond [9].

Experiments with impact disintegration of elastic solutions of high polymers have shown that elasticity radically changes the stability of droplets upon impact. The drops become much more durable [31]. During collision the secondary droplets form in the rim of the splash, which, however, at sufficient elasticity of the liquid, do not detach from the splash rim, but turn out to be connected with the rim by thin thinning filaments. Under the action of the elasticity of the filaments, the secondary droplets return back, merging with the liquid of the original droplet, and no free secondary droplets are formed (Fig. 1, *c*). The drop retains its continuity. Moreover, at sufficient elasticity, the rim of the splash expands in the form of a smooth liquid torus. It first expands and then collapses, merging into a single drop (Fig. 1, *d*).

Studies [31] have shown that, together with the already mentioned Weber number $We_i=\rho v_i^2 d_i/\gamma$, the defining parameters describing the transition to the disintegration of an elastic droplet upon collision are the elastic-capillary number $Gd_i/\gamma$ and the Deborah number $De_i=\theta v_i/d_i$, where $G$ is the modulus of elasticity, $\theta$ is the relaxation time. Rheological parameters $G$ and $\theta$ are constants of Oldroyd constitutive equation (Oldroyd-B version) and Maxwell (Upper-Convected Maxwell UCM version) [2, 12, 19]. Both equations turn out to be equivalent for describing rapid elongation deformations during strong stretching (see the Theoretical model section).

Another aspect of the influence of fluid elasticity on the transmission of infections is the possibility of elasticity to influence the relatively slow movement of fluid through channels of complex shapes or a porous medium [11, 13], which occurs when different types of filters are used for the protection from infected droplets. The same parameters of Oldroyd (Oldroyd-B version) and Maxwell (Upper-Convected Maxwell UCM version), namely $G$ and $\theta$, control elastic effects in these classes of flows.

The present work was undertaken to answer the question: what are the magnitudes of the rheological parameters of the oral fluid $G$ and $\theta$ that can be used to simulate the collision of oral fluid droplets with different obstacles or during the motion through complex media.

Until now, rheological tests of the oral fluid were carried out only to determine solely one parameter, namely, the relaxation time $\theta$ [10, 18, 41, 44]. At the same time, in order to have a complete rheological description of the properties of the oral fluid, it is necessary to establish the magnitudes of the elastic modulus $G$. The pair $G$ and $\theta$ makes it possible to model not only the disintegration of self-thinning filaments, but also model other phenomena in which the oral fluid participates, such as the disintegration of drops upon collision with obstacles and movement through filters and channels of complex shapes.

## EXPERIMENTAL OBSERVATIONS

This section is devoted to the method of rheological testing of an oral fluid using a thinning capillary filament of this fluid [6]. Experiments on observation of the breakup of oral fluid droplets *in situ* have been carried out. Oral fluid is an aqueous solution of various organic and mineral components, the total concentration of which is *ca*. 0.5% [27]. Among the main components is a mixture of secretions of the salivary glands. Among the glands, large ones are distinguished: parotid, submandibular, sublingual and a number of small salivary glands (1 - 5 mm in size) - buccal, molar, labial, lingual hard and soft palate. The names themselves indicate a variety of sources of salivary secretions and, therefore, about the possible heterogeneity of the mixture formed. The density of the oral fluid practically does not differ from the density of water $\rho = 1001 - 1017$ kg/m$^3$.

The oral fluid was sampled with tweezers directly from the lips of the subject. At the present stage, the author has conducted experiments on himself (a 65-year-old man). The simplest experimental procedure was used. Immediately after taking a sample of the oral fluid, it was stretched between the legs of the forceps. Further, the formed liquid filament thinned by itself without any external interference. The process was recorded using a video camera of an iPhone 6 smartphone at 30 frames per second. The possible 240 frames per second mode did not reveal any advantages and was rarely used in experiments. The frame resolution used was Full HD 1920×1080. To obtain dimensional dependencies, a transparent ruler was placed in the record field. A photograph of the equipment used is shown in Fig. 2.

It should be noted that the use of a smartphone to visualize the thinning of capillary filaments of polymer liquids was apparently first demonstrated in [22], which showed the possibility of a significant reduction in the cost of this experimental technique.

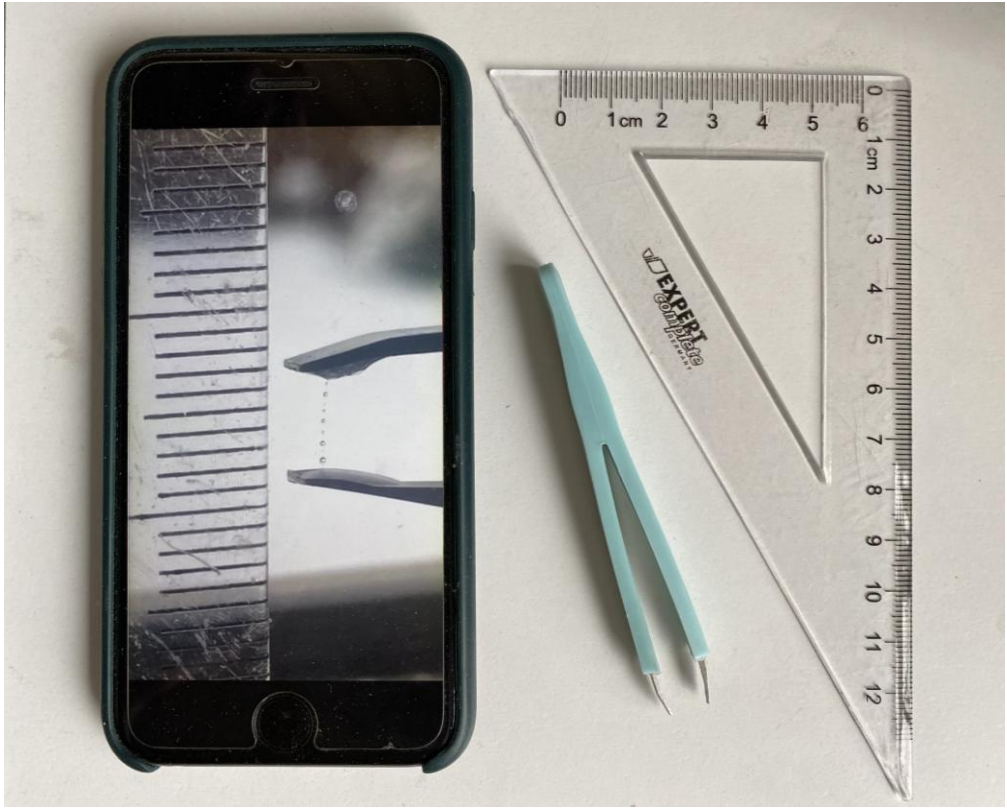

Fig. 2. Experimental equipment for studying the rheology of elastic fluids: a smartphone, tweezers and a ruler

A typical video recording is shown in Fig. 3. Video data was analyzed using the QuickTime Player application, although other applications such as, for example, ImageJ or Phantom can be used also.

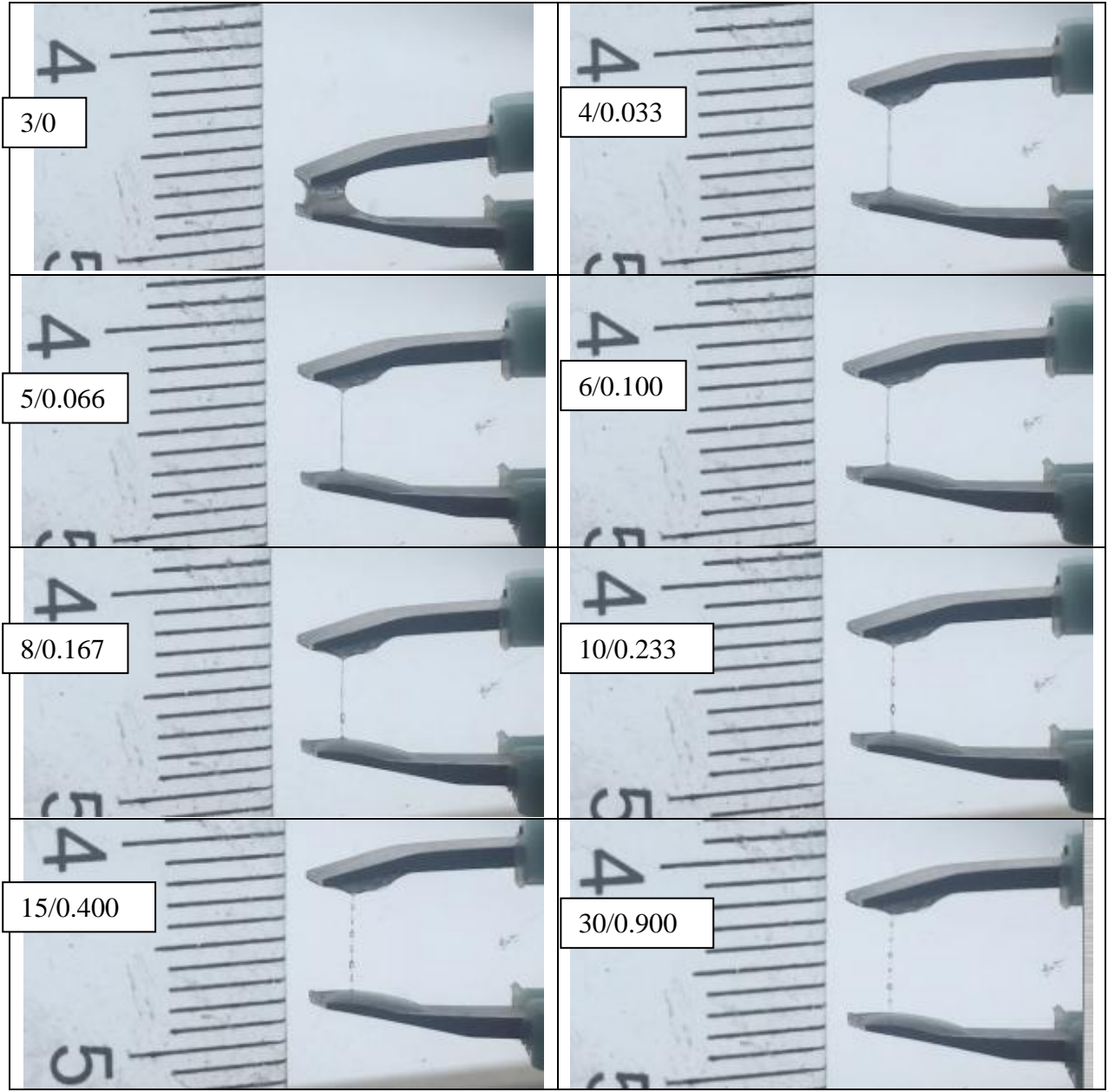

Fig. 3. Video recording of the formation and thinning of the oral fluid filament (experiment 7114, also see Figs. 4, 5). Time increases from left to right and from top to bottom. The insets indicate the frame number / filament thinning time (seconds)

The filament becomes thinner over time. At some stage, secondary breakup is possible, when a series of secondary droplets appears. Measurements of the filament diameter on video frames made it possible to plot the dependence of the thinning of the oral fluid filament in time $d = d(t)$. The appearance of secondary droplets at the later stages of breakup was ignored in the measurements. The start of the countdown was selected at the moment the tweezers' legs were extended (frame 3). The scale of the images was determined using a transparent ruler. Fig. 4 shows a typical dependence. The data indicate that immediately after the start of stretching, the liquid drop loses its stability and rapidly becomes thinner under the action of surface tension from the initial diameter $d_b$ to some intermediate diameter $d_s$ at a certain moment of time $t_s$. This phase of deformation corresponds to the stretching of a drop of an Newtonian fluid [5]. At this stage of stretching, the surface tension forces are balanced by the inertial forces of the fluid and internal stresses. Internal stresses are initially of a viscous nature, but as the thinning progresses, elastic stresses appear due to the rapid deformation of the liquid. This is how high-polymer solutions behave, in which the solvent plays the role of a viscous liquid, and elasticity is formed by giant stretching of elastic flexible-chain of macromolecules [5]. At a certain moment (qualitatively denoted by $t_s$), the elasticity begins to dominate over all other forces and the process of filament disintegration slows down sharply. Further thinning of the filament becomes exponential, as evidenced by the data in Fig. 4.

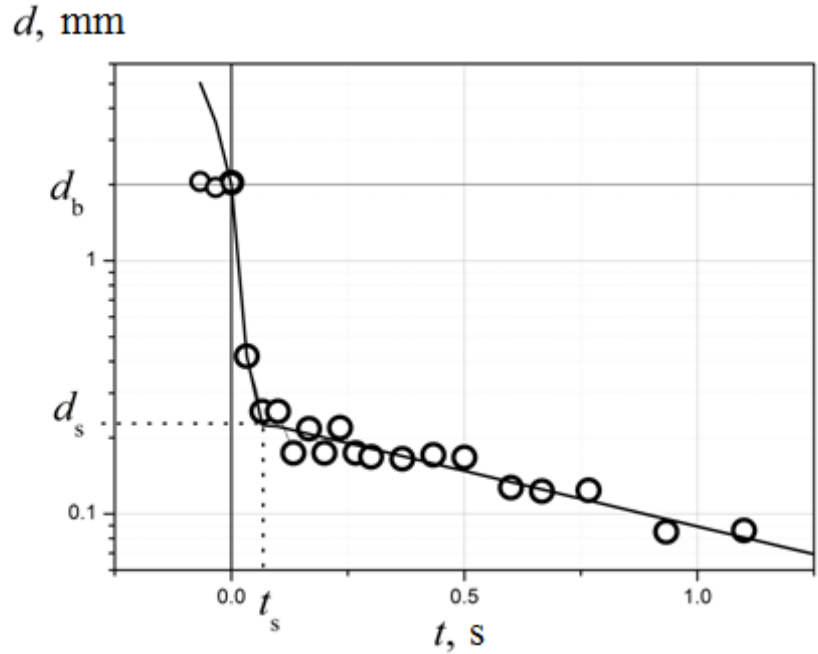

Fig. 4. Thinning of the filament of the oral fluid − dependence of the filament diameter as function of time. Solid curve − fitting of experimental points (experiment 7114) by analytical dependence (6) for points $t>0$

## THEORETICAL MODEL

This section presents a technique for extracting the rheological characteristics of an oral fluid from observations of the thinning of a filament of this fluid.

Studies of the rheological properties of the oral fluid under tension using the thinning fluid method [10] have shown that the rheological behavior of the oral fluid under strong stretching is similar to the behavior of polymer solutions [5]. To describe the rheology of polymer solutions under strong stretching, the Oldroyd constitutive equation (Oldroyd-B version) [12, 19] and Maxwell constitutive equation (Upper-Convected Maxwell UCM version) [2, 12, 19] were used.

Oldroyd constitutive equation (Oldroyd-B variant) is described by the equations

$$\boldsymbol{\sigma} = -p\boldsymbol{I} + 2\mu_s \boldsymbol{E} + \boldsymbol{\tau}, \quad \boldsymbol{\tau} = G(\boldsymbol{A}-1),$$

$$d\boldsymbol{A}/dt = \boldsymbol{A}\cdot\nabla\boldsymbol{v} + \nabla\boldsymbol{v}^{\mathrm{T}}\cdot\boldsymbol{A} - (\boldsymbol{A}-\boldsymbol{I})/\theta,$$

where $\boldsymbol{\sigma}$ is the stress tensor, $p$ is the pressure, $\mu_s$ is the solvent viscosity, $G$ is the elastic modulus, $\boldsymbol{E}$ is the strain rate tensor, $\boldsymbol{A}$ is the elastic strain tensor, $\boldsymbol{I}$ is the unit tensor, $t$ is the time, $\nabla\boldsymbol{v}$ is the velocity gradient, $\theta$ is the relaxation time.

Considering the elongation flow in direction 1 with the accumulation of large elastic deformations ($A_1 \gg 1$), we obtain the dependence of the stresses on the elongation kinematics

$$A_1 = \lambda^2 \exp(-t/\theta), \quad \tau_1 = GA_1, \tag{1}$$

where $\lambda$ is the elongation of the specimen in the direction 1, $\lambda \gg 1$.

Maxwell constitutive equation (Upper-Convected Maxwell UCM version) is represented by equations

$$\boldsymbol{\sigma} = -p\boldsymbol{I} + \boldsymbol{\tau},$$

$$d\boldsymbol{\tau}/dt = \boldsymbol{\tau}\cdot\boldsymbol{E} + \boldsymbol{\tau}\cdot\boldsymbol{E} - \boldsymbol{\tau}/\theta + 2\mu\boldsymbol{E}/\theta,$$

where $\boldsymbol{\sigma}$ is the stress tensor, $\boldsymbol{\tau}$ is the deviatoric stress tensor, $p$ is the pressure, $\boldsymbol{I}$ is the unit tensor, $\mu$ is the elongation viscosity, $\boldsymbol{E}$ is the strain rate tensor, $\theta$ is the relaxation time, $t$ is the time.

In the case of large elastic deformations ($\lambda^2 \gg 1$, $\tau_1 \gg \mu/\theta$) this implies

$$\tau_1 = (\mu/\theta)\lambda^2 \exp(-t/\theta), \qquad (2)$$

where for elastic polymer liquids the ratio $\mu/\theta$ is the elastic modulus $G$ [43]. Thus, in the case of dominance of elastic stresses, the axial component of the elastic stress tensor (1) and the axial component of the deviatoric stress tensor (2) coincide, which indicates the equivalence of the Oldroyd and Maxwell constitutive equations for describing the flows of elastic fluids with large deformations. The mechanical analogue of this rheological constitutive equation is a serial connection of viscous and elastic elements [21] – Fig. 5, *a*.

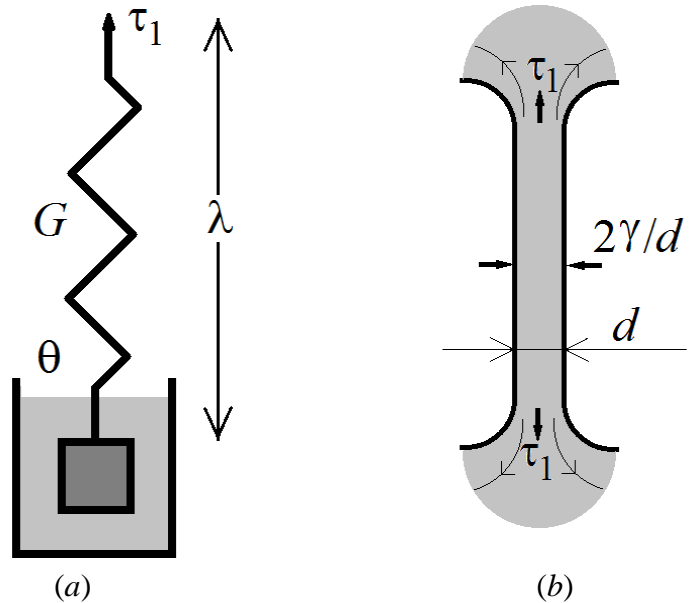

Fig. 5. Models of viscoelastic rheology (*a*) and thinning filament (*b*)

As shown by direct experimental measurements of the elastic tension of the filament of a high-polymer solution at the stage of exponential thinning ($t > t_s$) [4, 7], the elastic component $\tau_1$ approximately coincides with the value of the capillary pressure in the filament $\tau_1 \approx 2\gamma/d$ (Fig. 5, b), where $\gamma$ is the surface tension. Assuming for the degree of elongation $\lambda = (d_b/d)^2$, from (1) (or (2)) we obtain the formula for the filament thinning at the exponential stage

$$2\gamma/d = G(d_b/d)^4 \exp(-t/\theta),$$

which can be represented as

$$d = d_s \exp(-(t-t_s)/3\theta), \qquad (3)$$

where the variables of equation (3) are related to each other by the condition

$$(2\gamma/G)(d_s^3/d_b^4)\exp(t_s/\theta) = 1. \qquad (4)$$

The initial fast stage of filament thinning $t < t_s$ was modeled by a linear dependence with the proportionality coefficient α

$$d = d_b - \alpha t, \qquad (5)$$

which qualitatively corresponds to the thinning of the capillary filament of a viscous liquid [5, 6, 9, 10, 42]. In this work, formula (5) was used only for kinematic estimates.

The rheological parameters of the liquid were determined by the fitting the experimental dependence $d=d(t)$ (see, for example, Fig. 4) by an analytical dependence constructed on the basis of the superposition of theoretical curves (3) and (5).

$$d = (d_b - \alpha t)H(t_s - t) + d_s \exp(-(t - t_s)/3\theta)H(t - t_s), \qquad (6)$$

where $H$ is the Heaviside function, $d_s = d_b - \alpha t_s$ due to the continuity of the theoretical dependence $d=d(t)$. This curve is continuous and consists of two branches, the first of which describes the fast initial stage of filament thinning (5) at $t < t_s$, and the second, slow exponential (3) at $t > t_s$. Joining of solutions at the point $t_s$, $d_s$ determines the boundary of the transition from one flow regime to another.

The fitting was carried out using the Origin 6.1 application (Section: Analysis, NonLinear Curve Fitting). The parameters α, $t_s$, θ were to be determined. In the calculations, the fitting function was specified as (for experiment 7114)

$$(2.0-\text{ALF}*x)/(1+\exp(-2*100*(-x+ts))) + (2.0-\text{ALF}*ts)*\exp(-(x-ts)/3/\text{TETA})/(1+\exp(-2*100*(x-ts))),$$

in which $2.0 = d_b$ is the parameter $d_b$ measured in this experiment (see Fig. 3 and 4); ALF, ts, TETA are the sought parameters α, $t_s$, θ, respectively; x is the independent variable corresponding to time $t$; $1/(1+\exp(-2*100*(-x+ts)))$ and $1/(1+\exp(-2*100*(x-ts)))$ are smooth modelings of Heaviside functions $H(t_s - t)$ and $H(t - t_s)$. As a result of numerical calculations, the unknown parameters were found and fitting curve was constructed, see the example in Fig. 4.

Video records of 8 experiments was processed, the results are presented in the graph in Fig. 6 and Table 1. The relaxation time θ was determined directly as a result of the fitting of the experimental points, and the elastic modulus $G$ was calculated by the formula (4). In the calculations, the surface tension of the oral fluid was assumed to be $\gamma = 0.050$ N/m [25, 39].

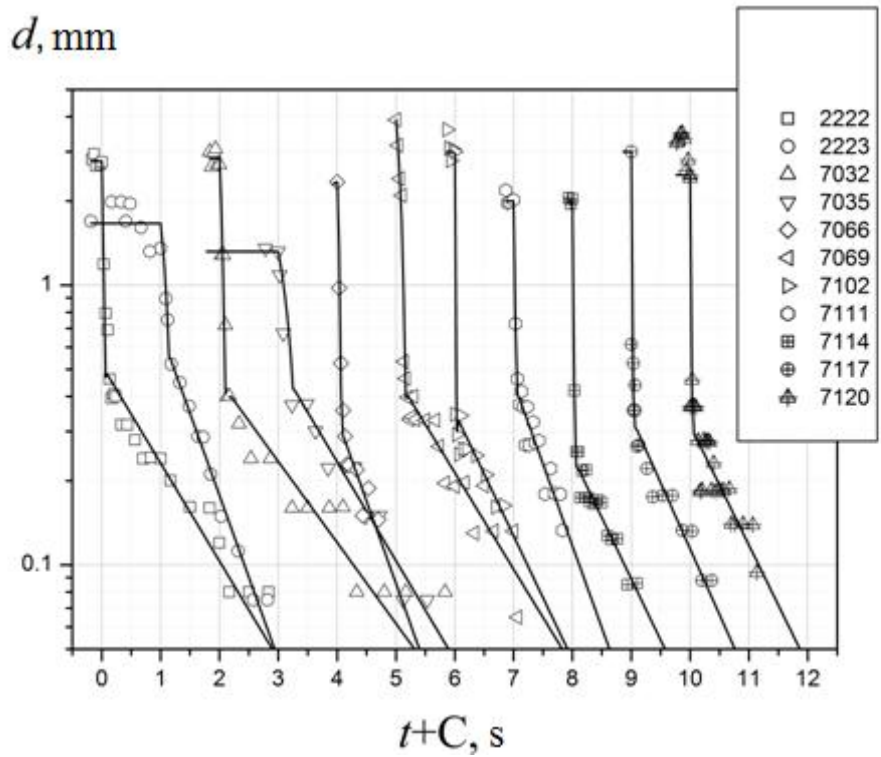

Fig. 6. Dependences of the filament diameters versus time according to the data of 8 experiments (symbols and numbers of experiments are shown in the inset). For clarity, the curves are shifted in time with a step of 1 s. The horizontal lines in some curves show the droplet size before the start of stretching. Solid piecewise linear dependences are the fittings of experimental points by analytical dependence (6)

*Table 1*

**Data processing results**

| Experiment, № | $d_b$, mm | $d_s$, mm | $t_s$, s | $G$, Pa | $\theta$, s |
|---|---|---|---|---|---|
| 2222 | 2.777 | 0.4977 | 0.04795 | 0.2328 | 0.41364 |
| 2223 | 1.6627 | 0.5658 | 0.14098 | 4.1957 | 0.24684 |
| 7032 | 2.8485 | 0.4308 | 0.08505 | 0.1440 | 0.49996 |
| 7035 | 1.319 | 0.4320 | 0.24703 | 4.8745 | 0.40877 |
| 7066 | 2.336 | 0.3022 | 0.07169 | 0.1241 | 0.24563 |
| 7069 | 3.9 | 0.4182 | 0.1533 | 0.0856 | 0.41975 |
| 7102 | 3.0 | 0.3448 | 0.0199 | 0.0538 | 0.32747 |
| 7111 | 2.0 | 0.4257 | 0.04266 | 0.5730 | 0.24713 |
| 7114 | 2.0 | 0.2353 | 0.03873 | 0.0916 | 0.33019 |
| 7117 | 3.0 | 0.3252 | 0.03535 | 0.0476 | 0.3073 |
| 7120 | 2.477 | 0.3047 | 0.03585 | 0.0836 | 0.33761 |

The measured rheological data (Table 1) are characterized by an extremely high scatter, although a quick look at Fig. 5 does not give a priori grounds to assume such a feature. Two reasons contribute to the scatter in the output, especially noticeable for the modulus of elasticity. First, as noted earlier, the oral fluid is formed by mixing the secretions of several salivary glands, each of which can form a secret with unique properties. Another reason is the strong dependence on input data. In particular the modulus of elasticity is determined by raising the magnitude of intermediate diameter $d_s$ to the third power and the magnitude of initial diameter $d_b$ to the fourth power. Therefore, a small error in $d_s$, $d_b$ values is amplified during the exponentiation process. This situation is close to the case of an ill-posed problem [38]. Taking into account the above circumstances, it is advisable to use the medians as data averaging [23]. "Estimates of the median are more robust, its estimation may be more preferable for distributions with the so-called heavy tails. The main advantage of the median is its robustness to outliers, i.e. abnormally high or low values." Using the Origin 6.1 application to analyze the data in Table 1 leads to the following median values of the rheological parameters of the oral fluid

$$<G>=0.10785 \text{ Pa}, <\theta>=0.32883 \text{ s}. \tag{7}$$

Comparing the parameters (7) with the results of measurements of similar parameters in aqueous solutions of polyacrylamide (PAM) and polyethylene oxide (PEO) [31], one can conclude that the elastic parameter $<G>=0.10785$ Pa for the oral fluid is close to elastic parameters of solutions of PAM and PEO with concentrations of 252 and 12449 ppm. The oral fluid relaxation time $<\theta>=0.32883$ s is close to similar data of solutions of PAM and PEO with concentrations of 2102 and 19320 ppm, respectively (Fig. 7). This means that type of polymers and the appropriate concentration of solution can be selected to prepare polymer solutions that simulate the rheology of real oral fluid. In particular, a PAM solution with a concentration of 200 - 2000 ppm (0.02 - 0.2%) and a PEO solution with a concentration of 10000 – 20000 ppm (1 - 2%) are quite close in their rheology to a real oral fluid.

In [17, 32, 33], PAA solutions with concentrations of 100 and 1000 ppm were used to simulate the collision of a droplet of oral fluid with a thin thread and a plate with a hole (as elements of a medical mask and air filter).

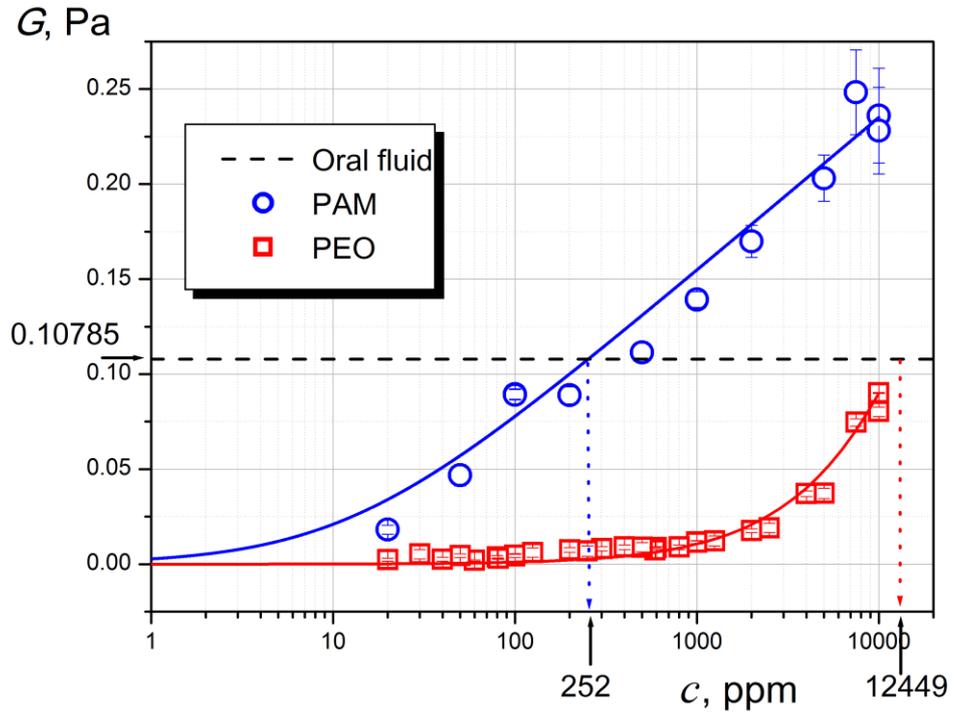

(a)

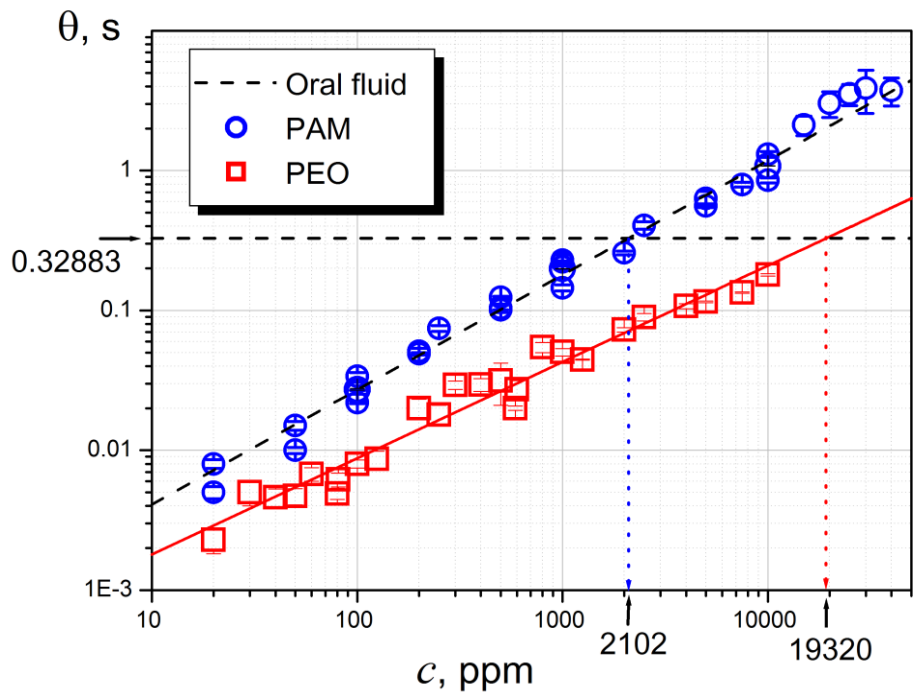

(b)

Fig. 7. Comparison of the measured rheological characteristics of the oral fluid (elastic modulus $G$ (a) and relaxation time $\theta$ (b)) with the data of similar measurements for aqueous solutions of PAM and PEO of various concentrations $c$ [31].

## CONCLUSIONS

Important consequences arise from the work performed. First, the full set of numerical values of the rheological characteristics of the oral fluid, namely, the elastic modulus $G$ and the relaxation time $\theta$, was determined/evaluated for the first time. (The previous measurements concerned only the relaxation time.) Establishing the physical characteristics of the oral fluid makes it possible to choose among the possible polymer solutions those whose rheology coincides with the rheology of a real oral fluid. The use of the selected model polymer fluids makes it possible to experimentally reproduce in the laboratory hydrodynamic processes occurring with a real oral fluid.

Second, the set of $G$ and $\theta$ "closes" the rheological equations of state of Oldroyd-B or Maxwell with the upper convective derivative, which describe the fast stretching of viscoelastic fluids. It is these processes that take place in the oral fluid during the ejection of drops during coughing, sneezing, talking. The further fate of the drops, namely their interaction (including coalescence and splashing) with the material of protective masks, filters and other obstacles, can also be described using these models. In particular, determined physical data will help predict what will happen with the drops in the surrounding space, i.e. will drops fall on the floor, consolidate in protective equipment, or collapse to the level of an aerosol which become convenient objects for the transfer of pathogenic agents by convective air currents? The description of the interaction of viscoelastic droplets with obstacles can be useful in the development of new methods of protection against infections.

Finally, using the oral fluid as an example, the possibility of rheological tests of biological elastic fluids is shown using the simplest (in fact, household) instruments - a ruler, a smartphone, and tweezers. The progress in consumer electronics have brought this class of technology closer to the capabilities of professional scientific instruments. The procedure for the determining the rheological characteristics of fluids based on video processing and numerical analysis of the resulting dependencies is demonstrated. In general, the algorithm for performing the presented rheological test is extremely simple and accessible to any researcher. In fact, the test replaces rheological tests carried out using the relatively expensive CaBER1 instrument [36]. The technique is applicable to studies of the rheological properties of biological fluids, which can form noticeable thinning capillary filaments. This class of fluids includes, for example, sputum [3, 8, 15, 28], synovial fluid [34, 35], fluids involved in the reproductive process (see, for example, [37]), and others. All of them can be investigated using the proposed method. The presented algorithm for biorheological research can be implemented in a special smartphone application for use in a variety of conditions, not necessarily laboratory ones.

## ACKNOWLEDGEMENTS

This work was carried out within the framework of State Program No. AAAA-A17-117021310375-7 and with the support of the RFBR grant 20-04-60128 Viruses.13